# Multi-Connectivity in 5G terrestrial-Satellite Networks: the 5G-ALLSTAR Solution

F. Lisi[1], G. Losquadro[2], A. Tortorelli[1], A. Ornatelli[1], M. Donsante[2]

[1] Sapienza University of Rome, via Ariosto 25, 00185 Rome, Italy, +39 0677274037,
{lisi, dellipriscoli, tortorelli, ornatelli}@diag.uniroma1.it

[2] CRAT, Via Giovanni Nicotera 29, Italy, +39 0677274037, giacinto.losquadro@crat.eu
manueldonsante@gmail.com

## Abstract

The 5G-ALLSTAR project is aimed at integrating Terrestrial and Satellite Networks for satisfying the highly challenging and demanding requirements of the 5G use cases. The integration of the two networks is a key feature to assure the service continuity in challenging communication situations (e.g., emergency cases, marine, railway, etc..) by avoiding service interruptions. The 5G NTN (Non-Terrestrial Network e.g., Satellite Network) would have a fundamental role in 5G [1], thanks to the its characteristics exploited typically for live events broadcasting in large areas and for ultra-reliable and secure communications; The networks integration will have a great impact to the network performances. The 5G-ALLSTAR project proposes to develop Multi-Connectivity (MC) solutions in order to guarantee network reliability and improve the throughput and latency for each connection between User Equipment (UE) and network. In the 5G-ALLSTAR vision to easily integrate the terrestrial and satellite networks, allowing Fast Switching and User Plane Aggregation, we divide the gNB in two entities [3]: 1) gNB-CU (Centralized Unit) and 2) gNB-DU (Distributed Unit). Each gNB-CU controls a set of different gNB-DUs (see Figure 1-a). The gNB-CU integrates an innovative Traffic Flow Control algorithm able to optimize the network resources by coordinating the controlled gNB-DUs resources, while implementing MC solutions. The MC [2] permits to connect each UE (whether possible) with simultaneous multiple access points which can belong to both the same and different radio access technologies. The 5G-ALLSTAR solution for the MC deals with the possibility to have a common RRC and partial User Plane functionalities in the gNB-CU, i.e., SDAP and PDCP layers. This solution leads to have independent gNB-DU/s that contain the RLC, MAC and PHY layers. The communication between gNB-CU and the controlled gNB-DUs takes place by using the well-known F1 interface [3]. As an example of integration between NTN and Terrestrial Networks we can consider the MC solution shown in Figure 1-b where the same packet (duplicated by the PDCP layer) are delivered independently to the two access points (gNB-DU-SAT and gNB-DU-5G) [4],[5],[6]. The 5G-ALLSTAR MC algorithms offer advanced functionalities to RRC layer [7] (in the gNB-CU) that is, in turn, able to set up the SDAP [8], the PDCP [9] and the lower layers in gNB-DU. In this regard, the AI-based MC algorithms, implemented in gNB-CU (also known as Cloud RAN), by considering the network performances in the UE surrounding environment as well as the UE QoS requirements, will dynamically select the most promising access points able to guarantee the fulfilment of the requirements (guarantying the required degree of throughput and latency) also enabling the optimal traffic splitting to cope with the connection reliability. In this paper, we present also an innovative AI-based framework, included within the Traffic Flow Control, able to address the MC objectives (as presented above), by implementing a Reinforcement Learning algorithm in charge of solving the network control problem.

## 1. Introduction

The Multi Radio Access Networks (Multi-RAT) or heterogeneous access networks, are believed to be one the key point of the new and highly challenging 5G requirements. The 5G requirements that could benefit from Multi-RAT are high data rate, ultra-low latency and network reliability. To address such requirements in an efficient way, the multi-connectivity techniques has been proposed to simultaneously link, and control, several different (at least one) radio access technologies to the same User Equipment (UE) at the same time. An important 5G requirement that would enable advanced solution in the IT solutions, network reliability, is not addressed when an established connection fails during the data transmission; in fact, to address network reliability we should consider the same UE linked with at least two different access points, so that in the case in which the connection with one of the two access points fails, the data transmission is still guaranteed. In order to enable the network reliability, the 5G-ALLSTAR project proposes a Multi-Connectivity solution which is able to overcome the issues introduced by Dual-Connectivity [2] and also easy integrate Terrestrial and Satellite Networks. The Multi Connectivity solution, as it has been proposed in 5G-ALLSTAR project, includes an architectural framework able to address the Multi-RAT connections, and an AI-based framework able to address the mentioned 5G requirements. In this work we will present the architectural framework in charge of allowing Fast Switching and User Plane Aggregation and we will introduce only the potential algorithm solutions that are currently under ingestion within the 5G-ALLSTAR project. The paper is organized as follow: in Section 2 we present the 5G-ALLSTAR project and the state of the art concerning the current MC solutions from architectural and algorithm point of view; in Section 3 we present the Dual-Connectivity, Multi-Carrier and Multi-Connectivity architecture; in Section 4 we present what

5G-ALLSTAR proposes to includes its solution within the current 5G system; Section 5 presents conclusion and future works.

## 2. 5G-ALLSTAR Project

The developing of new generation cellular network is a consequence of the necessity of a new communication network, able to provide a common platform supporting new services and seamless connectivity to various vertical industries and for very different use cases. 5G-ALLSTAR aims to design, develop and verify multi-connectivity solutions, based on multi-RATs, enabling the 5G network requirements. Namely, the project objective is the development of novel multi-connectivity architecture and techniques able to allow automatic, agile and flexible handover between the involved heterogeneous technologies, considering the actual network state and the actual users and service's needs. The outputs of the project will be:
contributions to the 5G standardization on multi-connectivity;
- the development of mmWave cellular access system for broadband and low latency services;
- the study of new radio satellite access to provide broadband and reliable services;
- the development of a multi-connectivity support (i.e., multi-RATs access network architecture and control techniques) to integrate in a seamless way the two technologies;
- the development of spectrum sharing techniques between satellite and terrestrial access technologies.

Finally, a European-Korean interoperable proof-of-concept to demonstrate the above technologies in selected use cases will be developed.

## 3. State of the Art

Multi-Connectivity should be seen from a twofold perspective the architectural and the algorithmic point of view. Multi-Connectivity architecture: several architectures architectural [10]-[22], depending on the selected integration approach, were proposed. The multi-RATs integration can be performed at three main levels three main integration approaches [23]. The levels of integration are Application layer integration, Core-Network based integration and RAN base integration. The Application layer integration can be easier to develop and manage, but in the case of very dynamic changes in the network state, this method leads to suboptimal solutions, the Core-Network integration, adopted by 3GPP for cellular/WLAN integration based on interworking between core networks, allows the use of operators' policy for network selection, but the final decision is in control of the UEs, that has only local knowledge about the network conditions, and can apply selections that degrade the overall network conditions. The RAN based integration allows the cooperation between RATs in the same access network and allows a fast response to radio link conditions variation minimizing session interruption and packet drops, maintaining the overall access network conditions optimal, it can result hard to implement and manage. Therefore, the three possible approaches are User-Centric approach, RAN-Assisted approach, and RAN-Controlled approach. In the User-Centric approach, the RAT selection is based on the measurement provided by the UE, e.g. the SNR below a defined threshold activates a RAT handover, this solution can be suboptimal due to the poor attributes used for the selection. The RAN-Assisted approach has the added value with respect to the User-Centric considering RATs status information, e.g. RATs load. The RAN-Controlled approach, adopted by 3GPP for addressing dual-connectivity issues, places the multi-connectivity control in the access network, and the UEs are not able to make decisions, but they are configured to report radio measurements on their local radio environment. This solution is optimal due to the knowledge of the entire status of the network. The RAN-Based Integration and RAN-Controlled approaches have been adopted in 3GPP Dual-Connectivity [2] in Figure 1, in the Metis-II project [10] and also presented in the 5GPPP document [11].

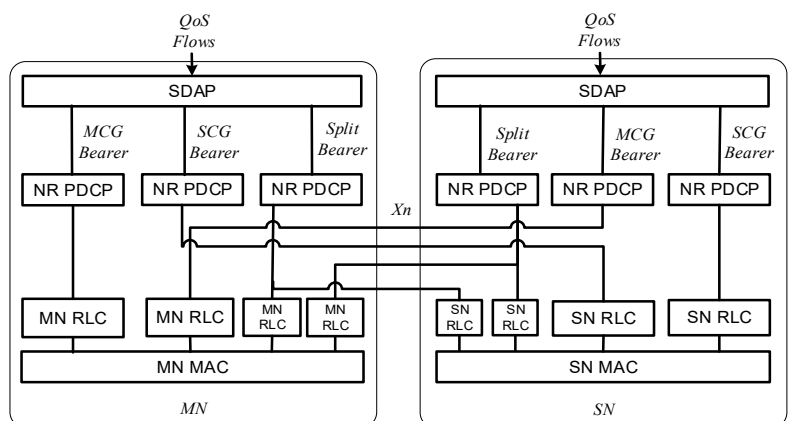

Figure 1: RAN side protocol termination options for MCG, SCG and split bearers in MR-DC with 5GC, [2]

The Metis-II project was aimed at designing an access network architecture and proving technical enablers for efficient integration between RATs. The project proposes an architecture as depicted in Figure 2 (b) composed by several Air Interface Variants (AIVs) and a central unit with AIV-agnostic functions, that based on the real-time feedback provided by the AIVs, is capable to steer the Quality of Service (QoS) Flows in a dynamic way on the different AIVs.

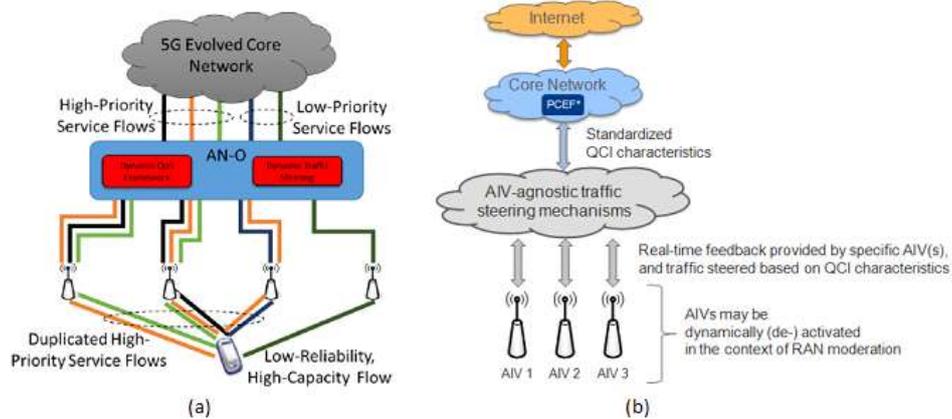

Figure 2: (a) Dynamic Traffic Steering framework in 5GPP [11]. (b) Architecture for traffic steering and RAN moderation in Metis-II [10]

Considering RAN-Based Integration and RAN-Controlled approaches presented above, the main topic of discussion is the functional split among RAN components. Following the 3GPP architecture in Figure 3, the gNB can be composed of a Central Unit (CU), also called either Central-RAN or Cloud-RAN and of several Distributed Units (DUs). This functional split between the Central and Distributed unit has the purpose of placing the cooperative, and technology independent, RAN functionalities in a central node, so that they can benefit from the advantages of centralization (i.e., centralized decisions, high computation power available), allowing flexible RATs selection/switch.

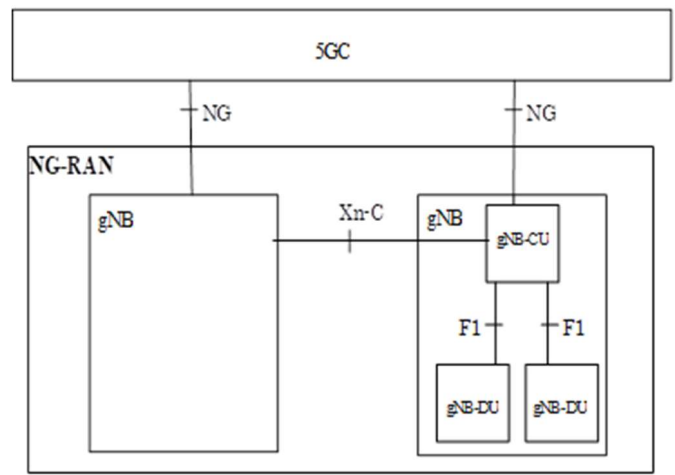

Figure 3: Overall RAN architecture [3]

The selection of the functional splits, i.e. the decision about which function should be placed in either the central or the distributed units, is a crucial point that defines the whole traffic flow control system.
From the control plane perspective, the ideal scenario would be to put the whole set of functionalities that are technology independent as well as Non-Real-Time and low bit rate, (e.g. traffic steering, spectrum sharing, etc, load balancing [24]…) in the central unit in order to have a complete view of the system, allowing optimal decision making. In this case, the distributed units have technology-dependent, Real-Time, and high bit rate functionalities in order to meet their requirements. Regarding the protocol stack split, there are three main options as shown in Figure 4, a suitable choice is a common PDCP for the user plane and a common RRC for the control plane. In contrast to PHY, MAC, and Radio Link Control (RLC) functions, the PDCP functions do not have rigorous constraints in terms of synchronicity with the lower layers. Furthermore, this option will allow traffic aggregation, as it can facilitate the management of traffic load. This split has already been standardized for LTE Dual Connectivity [25].

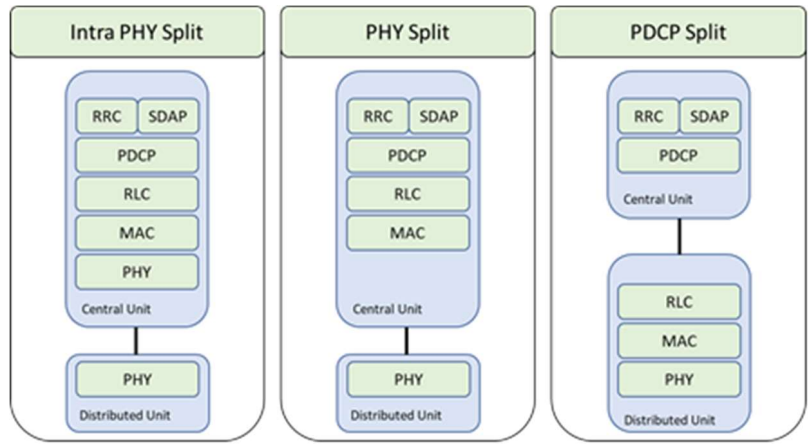

**Figure 4: Functional Split variants**

Multi-Connectivity algorithms: in [26], [27], MC is modeled as a traffic steering problem, the authors define the traffic steering as the function of distributing the traffic load optimally across different network entities and spectrum bands, considering operator (e.g., energy-saving, load optimization, interference management, and congestion management) and user preferences (e.g., QoE/QoS requirements). Furthermore, the traffic steering problem needs to consider the characteristics of the different RATs, such as coverage, latency and capacity and the different traffic characteristics, defined by the QoS requirements. Traffic steering can be performed in a centralized or distributed fashion. In both cases it allows the traffic steering algorithms to access information regarding load and serving capacity of all the RATs. Centralized coordination can achieve optimal performances, but this solution may introduce problems in the real-time control of the RATs. On the other hand, a distributed implementation can only perform traffic steering based on local information achieving suboptimal solutions, and the control traffic between RATs can be significant. In [28]-[30], MC is modeled as a RAT selection problem, also known as network selection problem in heterogeneous networks. The problem consists in the selection of the most appropriate access network with characteristics able to satisfy the 5G KPI requirements. These selections can be performed by considering different network features, e.g., the mobility of the network nodes, the QoS attributes, the energy constraints. The algorithms capable to perform the RAT selection are evaluated by considering the characteristics of the algorithms like computational complexity, implementation complexity, distributed or centralized deployment with either open or closed-loop type, dynamic or static behavior, model-based or data-driven. In Table 1 an overview of the mathematical theory used to solve such problem is presented.

**Table 1: Key Characteristics of Mathematical Theories for Network Selection**

|  | Utility Theory | MADM | Fuzzy Logic | Game Theory | Combinatorial Optimization | Markov Chain |
|---|---|---|---|---|---|---|
| **Objective** | Utility evaluation | Combination of multiple attributes | Imprecision handling | Equilibrium between multiple entities | Allocation of applications to networks | Consecutive decisions/ rank aggregation/ priority evaluation |
| **Decision Speed** | Fast | Fast | Fast | Middle | Slow | Middle |
| **Implementation Complexity** | Simple | Simple | Simple | Complex | Complex | Middle |
| **Precision** | Middle | High | Middle | High | High | High |
| **Model-Based/Data-Driven** | Model-Based | Model-Based | Model-Based | Model-Based | Model-Based | Data-Driven or Model-Based |
| **Open/Closed-loop** | Open-loop | Open-loop | Closed-loop or Open-loop | Closed-loop or Open-loop | Open-loop or Closed-loop | Closed-loop or Open-loop |
| **Centralized/Distributed** | Centralized | Centralized | Centralized | Centralized or Distributed | Centralized or Distributed | Centralized or Distributed |

4. **Multi-Connectivity Solution**

## Dual-Connectivity

In the 3GPP standard [2], a particular case of Multi-Connectivity is presented, i.e., the above-mentioned Multi-RAT Dual Connectivity, in which the multiple Tx/Rx UEs may be configured to use resources provided by two nodes: the first node provides E-UTRA access, while the second node provides NR access 5G ALL-STAR, the objective is to design, develop and test a more general multi-connectivity approach that is not only limited to NR and LTE but which also includes satellite and other terrestrial technologies such as WI-FI access. In 3GPP Dual-Connectivity, as reported in Figure 1, the two RATs involved in the connection are identified as Master and Secondary Nodes (MN and SN). Furthermore, three bearer types across the Uu interface are defined: i) the MCG bearers, only the MCG radio resources are involved; ii) the SCG bearers, only SCG radio resources are involved; iii) the Split bearers, both MCG and SCG radio resources are involved. Each SDAP entity, able to map the QoS Flow in the appropriate radio bearer/s, is placed in one RAT, but it works as a shared or "central" functionality. The MN decides which QoS flow should be assigned to each SDAP entity. The MN or SN node, that hosts the SDAP entity, for a given QoS flow decides how to map it to DRBs. The drawback in this approach is a large amount of communication between the MN and SN node, to configure the network entities and for the exchange of the data packets.

## Multi-Carrier (Carrier Aggregation)

In Carrier Aggregation (CA) [31], two or more Component Carriers (CCs) are aggregated. A UE may simultaneously receive or transmit on one or multiple CCs depending on its capabilities. In the case of CA, the multi-carrier nature of the physical layer is only exposed to the MAC layer for which one HARQ entity is required per serving cell as depicted in Figure 5 below:

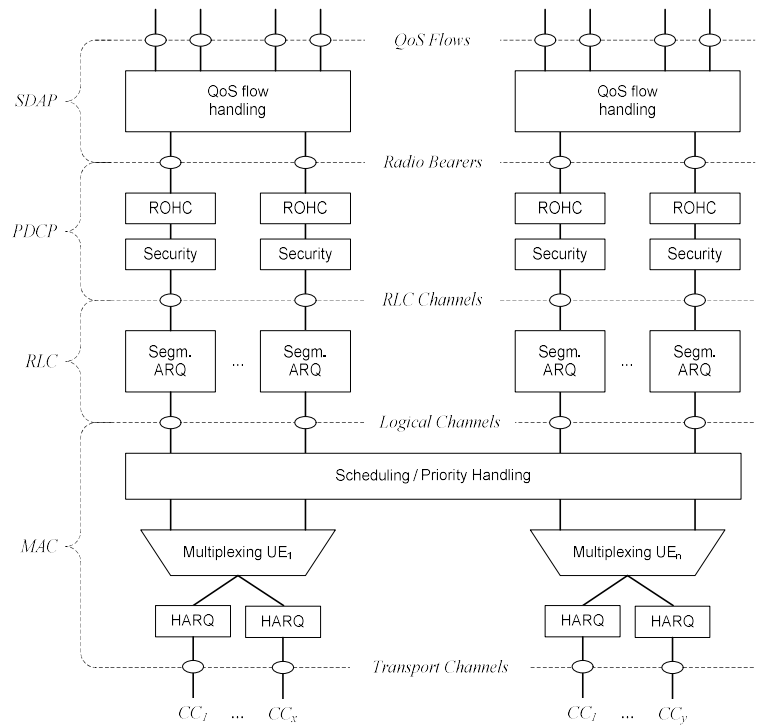

Figure 5: Layer 2 Structure for DL with CA configured [31]

In both uplink and downlink, there is one independent hybrid-ARQ entity per serving cell and one transport block is generated per assignment/grant per serving cell in the absence of spatial multiplexing. Each transport block and its potential HARQ retransmissions are mapped to a single serving cell. The disadvantage with respect to the proposed solution is the lower handover flexibility when CA is configured, the UE only has one RRC connection, this implies that in case of high mobility or link failure the handover procedure can cause packets drop.

## 5G-ALLSTAR Multi-RAT

The Multi-Connectivity Architecture, as it has been proposed in 5G-ALLSTAR project, is designed to be in line to the address the current 5G requirements and KPIs and also to be compliant with the distributed resource control. The architecture is presented in Figure 6. It is divided in 5G standard and 5G-ALLSTAR system to highlight how and where the 5G-ALLSTAR solutions will impact to the actual 5G Standard systems (which is currently considered in 3GPP [32]). In this regard, the 5G standard system is composed of two main entities, i.e., the 5G Core Network and the 5G Radio Access Network (RAN). In the 5G-ALLSTAR vision to easy integrate the terrestrial and satellite networks, allowing Fast Switching and User Plane Aggregation, we divide the gNB in two entities [3]: 1) gNB-CU (Centralized Unit) and 2) gNB-DU (Distributed Unit). Each gNB-CU controls a set of different gNB-DUs (see Figure 6). Note that each gNB-DU is considered to be a RAT. The gNB-CU integrates an innovative Traffic Flow Control algorithm able to optimize the network resources by coordinating the controlled gNB-DUs resources, while implementing MC solutions for an efficient centralization of traffic flow decisions (i.e., from both control and user-plane functionalities). The MC [2] permits to connect each UE (whether possible) with simultaneous multiple access points which can belong to both the same and different radio access technologies. The *standard 5G Core Network* and the *Standard 5G RAN* included in the Figure 6 contains the set of well know Core Network and RAN functionalities [32]. According to the 5G-ALLSTAR objectives the Core Network is enriched with Quality of Experience (QoE) and Quality of Service (QoS) functionalities (QoE/QoS Management [33]) able to provide/estimate a set of personalized parameters to guarantee the personalized degree of satisfaction. The gNB-CU in 5G-ALLSTAR is considered to be enriched with intelligent functionalities including a i) centralized Radio Resource Management (cRRM) able to provide Radio Link Performance; ii) the Traffic Flow Control, which is conceived to perform traffic switching, splitting and steering functionalities, provide control actions to the cRRM for enabling Multi-Connectivity. The gNB-DU in 5G-ALLSTAR is considered to be enriched with intelligent functionalities including a) Quality of Experience Estimation functionality is in charge of capturing for each on going connection between each RAT-UE couple the individual QoE (it has been described in [32]); b) a distributed Radio Resource Management (dRMM) which is in charge of monitoring the set of different cells performances. From algorithms point of view the 5G-ALLSTAR modules include both Machine Learning and control techniques based solutions.

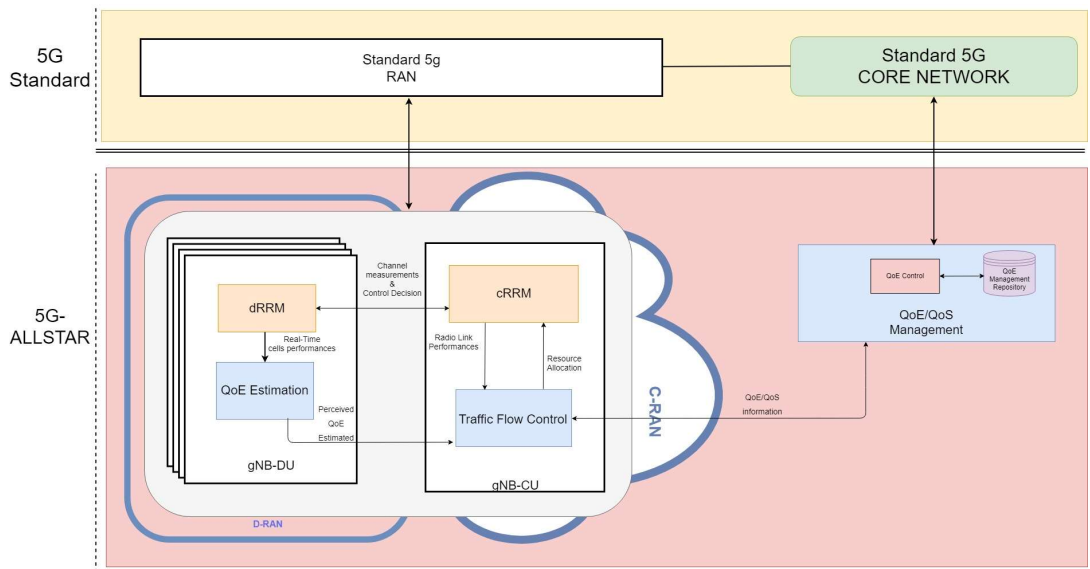

Figure 6: 5G-ALLSTAR Multi-Connectivity Functional Architecture (Control Plane) [32]

As already introduced the MC facilities involve multiple signal connections with different Radio Access Technologies, either considering the actual different access points (e.g., 5G, LTE, 3G, SAT, etc…) or different cells (for both macro and small cells) which belong to same access points. The simultaneous exploitation of multiple RATs, used for any single connection to serve a single UE (intended as multi-connectivity capable UE) can guarantee I) network reliability, and II) improve the whole data rate (throughput). When the MC is performed at the gNB-CU level, as already introduced above the whole set of control functionalities are centralized; this entails that the distributed units (gNB-DU) have no communications among them and in case one of them fails – during the on-going connection – the gNB-CU can provide an flexible and easy handover without data loss also considering both Terrestrial (e.g., 5G,4G, etc…) and non-Terrestrial Networks (e.g., Satellite Networks). In this respect, 5G-ALLSTAR project proposes to have SDAP and PDCP level within the gNB-CU and the RLC, MAC, PHY within the gNB-DU. This allows a Fast Switch among the different RATs to address the requirement as introduce in I) and II). The Figure 7 and Figure 8 represent an example of network reliability and throughput, respectively.

Figure 7 shows how to exploit different radio access technologies (or multiple cells of the same radio access technology) for transmitting the same data and addressing network reliability issues. The gNB-CU splits the same packets in two different directions. Figure 8 deals with an example of the traffic loads optimal distribution of the traffic loads, involving two RATs for an efficient multiple transmission for augmenting the overall data rate of a single connection. This would be particularly suited for use cases with a highly resource demand The MC solution implemented by the 5G-ALLSTAR Traffic Flow Control deals with the optimal resources distribution in use cases with highly challenging simultaneous demand (e.g., people at live events, swarm of sensors, etc…) or resource scarcity (e.g., disaster recovery [34], rural area connectivity, etc…).

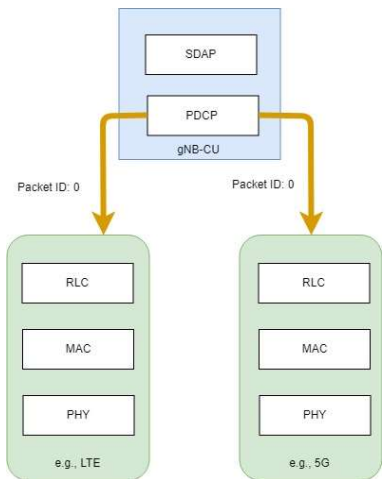
Figure 7: MC, network reliability [32]

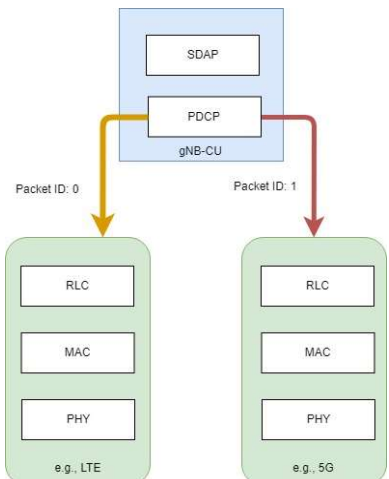
Figure 8: MC, increase the data rate [32]

## How Control the Network Functionalities

This section introduces the preliminary design of interfaces for (i) the inner 5G-ALLSTAR components and (ii) the communication between the 5G-ALLSTAR components and the standard 5G functional modules.
The 5G-ALLSTAR services expose public APIs in order to be triggered by external and internal functions. Such APIs (WP4 APIs) will be implemented and detailed during the project to foster the integration among the services and components implemented in the 5G-ALLSTAR work packages. In the following tables, the Interface/s Short Name is referred to the communication protocol among components as shown in Figure 9.

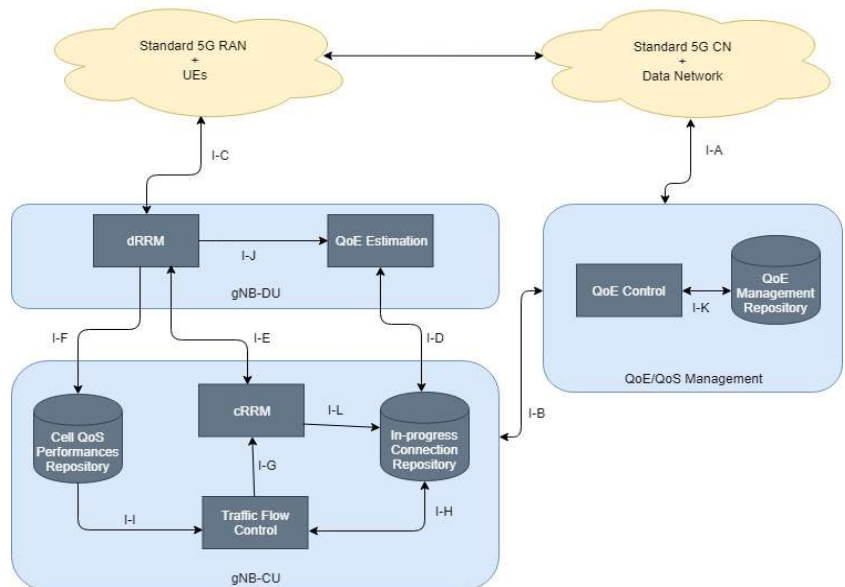
Figure 9: 5G-ALLSTAR Interfaces Diagram [32]

## 5. Conclusions and Future Work

Multi-RAT is the key enabling technology for next-generation communication networks, making a single network able to satisfy a wide range of services and users' requirements. Satellite technology is the best candidate to be integrated with the cellular network, thanks to its characteristics tank to its characteristics allowing live events broadcasting in large areas and ultra-reliable and secure communications, that are typically the weak points of the terrestrial networks. The proposed 5G-ALLSTAR solution allows a fast, flexible and context-aware integration and management of multi-RAT communications. The differences with respect to the actual solutions, such as Dual-Connectivity and Carrier Aggregation, are: i) the presence of the common control plane functionalities in central unit to manage the different RATs in coordinate and cooperative way, minimizing the communications between the different RATs; ii) the presence of the user plane higher layers in the central unit to aggregate/steer the traffic from/to the different RATs in agile and fast way, avoiding packets drop and data exchange between different network nodes; iii) the presence of Quality of Experience functionalities both in RAN and CN to configure the network considering the actual and past user experience; iv) the presence of a distributed RRM able to collect radio measurements for each RAT in an independent way. The proposed architecture can be easily integrated with the standard 5G network using external API and is highly modular to allow future developments. Future work will be conducted to develop and test Traffic Flow Control, QoE Management, and Radio Resources Management functionalities. The scope of the 5G-ALLSTAR project is to design and test multi-connectivity control solutions derived from the application of the mathematical tools mentioned in Table 1. The idea is to exploit Control Theory, Machine Learning/Artificial Intelligence, and Game Theory techniques to model, study and solve the multi-connectivity problem. The 5G-ALLSTAR next steps are to investigate and develop algorithms based on:
- Standard control methods for dynamical network control, e.g. [35];
- Reinforcement Learning based controllers, e.g. [36],[37],[38];
- Game-theoretic controls and equilibria, e.g. [39].

in compliance with the proposed architecture, and interfaces, of the previous sections.